\documentclass[12pt]{article}
\usepackage{cite}
\usepackage{url}
\usepackage{empheq}
\usepackage{staves}
\usepackage{amsthm,amsmath,latexsym,amssymb,amsfonts,amsbsy}
\usepackage{graphicx,lscape,fancyhdr,xcolor,array,stmaryrd,euscript,wrapfig}
\usepackage{hyperref}
\usepackage{mathrsfs}
\usepackage{cleveref}
\usepackage{physics} 
\usepackage{tikz-cd}
\usepackage{tikz}
\usepackage{tkz-graph}
\usetikzlibrary{topaths}
\usetikzlibrary{arrows}
\usepackage{mathdots}
\usetikzlibrary{positioning}
\usetikzlibrary{decorations.pathmorphing}
\definecolor{darkmagenta}{rgb}{0.55, 0.0, 0.55}
\definecolor{darkblue}{rgb}{0.0, 0.0, 0.55}
\definecolor{darkred}{rgb}{0.7, 0.0, 0.3}
\hypersetup{
	unicode,	% use with \texorpdfstring
	colorlinks,
	citecolor=darkred,% filecolor=black,% 
	linkcolor=darkblue, 
	urlcolor=darkred,% pdftex
	bookmarksopen=true,
	bookmarksnumbered
	}

 \csname
@addtoreset\endcsname{equation}{section}
\let\oldsqrt\sqrt
\def\sqrt{\mathpalette\DHLhksqrt}
\def\DHLhksqrt#1#2{%
\setbox0=\hbox{$#1\oldsqrt{#2\,}$}\dimen0=\ht0
\advance\dimen0-0.2\ht0
\setbox2=\hbox{\vrule height\ht0 depth -\dimen0}%
{\box0\lower0.9pt\box2}}

\begin{document}
\begin{titlepage}
\setcounter{page}{1}
\begin{center}
\hfill
\vskip -0.5cm

%~~~~~~~~~~~~~~~~~~~~~~~~~~~~~~~~~~~~~~~~~~~~~~~~~~~~~~~~~~~~~~~~
{\LARGE Liouville description of conical defects in dS$_4$,}\\[2mm] 
{\LARGE Gibbons--Hawking entropy as modular entropy, and }\\[2mm] 
{\LARGE dS$_3$ holography }
%~~~~~~~~~~~~~~~~~~~~~~~~~~~~~~~~~~~~~~~~~~~~~~~~~~~~~~~~~~~~~~~
\vskip 20pt
{\sc Cesar Arias}$^{a,b,}$\footnote{ \href{mailto:carias@math.ucdavis.edu}{\texttt{carias@math.ucdavis.edu}}},\, 
{\sc Felipe Diaz}$^{c,}$\footnote{ \href{mailto:f.diazmartinez@uandresbello.edu}{\texttt{f.diazmartinez@uandresbello.edu}}},\, 
{\sc Rodrigo Olea}$^{c,}$\footnote{ \href{mailto:mailto:rodrigo.olea@unab.cl}{\texttt{rodrigo.olea@unab.cl}}} 
~\&~
{\sc Per Sundell}$^{c,d,}$\footnote{ \href{mailto:per.anders.sundell@gmail.com}{\texttt{per.anders.sundell@gmail.com}}}
\vskip 20pt
${}^a${\em Department of Mathematics, University of California, Davis CA 95616, USA}  \\[1mm] 
${}^b${\em Riemann Center for Geometry and Physics, Leibniz 
Universit{\"a}t Hannover\\[-1mm]Appelstra{\ss}e 2, 30167 Hannover, 
Germany }\\[1mm]
${}^{c}${\em Departamento de Ciencias F\'isicas,  
Universidad Andres Bello\\[-1mm] 
Sazi\'e 2212, Piso 7, Santiago de Chile}\\[1mm]
${}^d${\em Department of Physics, Nanjing University, 22 Hankou Road, Nanjing, China} 

\vskip 20pt
{\bf Abstract}
\end{center}
\vskip -10pt
We model the back-reaction of a static observer in four-dimensional de Sitter spacetime by means of a singular~$\mathbb Z_q$ quotient.  
The set of fixed points of the~$\mathbb Z_q$ action consists of a pair of codimension two minimal surfaces given by 2-spheres in the Euclidean geometry.
The introduction of an orbifold parameter~$q>1$ permits the construction of an effective action for the bulk gravity theory with support on each of these minimal surfaces. 
The effective action corresponds to that of Liouville field theory on a 2-sphere with a finite vacuum expectation value of the Liouville field.    
The intrinsic Liouville theory description yields a thermal Cardy entropy that we reintrepret as a modular free energy at temperature $T=q^{-1}$, whereupon the Gibbons--Hawking entropy arises as the corresponding modular entropy.
We further observe that in the limit $q\to\infty$ the four-dimensional geometry reduces to that of global~dS$_3$ spacetime, where the two original minimal surfaces can be mapped to the future and past infinities of~dS$_3$ by means of a double Wick rotation. 
In this limit, the Liouville theories on the minimal surfaces become boundary theories at zero temperature whose total central charge equals that computed using the dS$_3$/CFT$_2$ correspondence.
{}\\
%
%%{\noindent \bf Abstract:} 
%\vspace*{4cm}
%\begin{flushleft}
%\footnotesize
%%\small 
%{$^a$\href{mailto:cesar.arias@unab.cl}{\tt cesar.arias@unab.cl}}\\
%{$^b$\href{mailto:f.diamartinez@uandresbello.edu}{\tt f.diazmartinez@uandresbello.edu}}\\
%{$^c$\href{mailto:rolea@unab.cl}{\tt rodrigo.olea@unab.cl}}\\
%{$^d$\href{mailto:per.anders.sundell@gmail.com}{\tt per.sundell@unab.cl}}
%\end{flushleft}
%%

\end{titlepage}

%%%%%%%%%%%%%%%%%%%
{\small
\tableofcontents }       
\vspace{1 cm}

%~~~~~~~~~~~~~~~~~~~~~~~~~~~
\section{Introduction}\label{Sec:Intro}
%~~~~~~~~~~~~~~~~~~~~~~~~~~~
The non-trivial topology of de Sitter (dS) spacetime comprises two disconnected spacelike boundaries and causally disconnected interior regions. Recently, it has been argued~\cite{Arias:2019pzy} that the Gibbons--Hawking entropy of dS spacetime \cite{Gibbons:1977mu} arises from the entanglement between the past and future conformal infinities or, alternatively, from the entanglement between two antipodal and causally disconnected bulk observers located at opposites Rindler wedges of the dS interior.

One of the central ideas behind the above argument is that in order to measure any observer-dependent quantity, in particular the thermal properties of the dS cosmological horizon, one has to go beyond the standard probe approximation of a static observer. In other words, the observer back-reaction should be taken into account.

Following this idea and motivated by the conically singular geometries induced by point particles in three dimensions~\cite{Deser:1983nh, Deser:1983tn, Deser:1989cf}, we model the back-reaction of a static observer in ${\rm dS}_4$ spacetime via the quotient ${\rm dS}_4/\mathbb Z_q$.  That is, we think of observers that back-react with the background geometry as inducing codimesion two defects that correspond to the fixed points of the $\mathbb Z_q$ action. In this sense, we treat the orbifold ${\rm dS}_4/\mathbb Z_q$ as the fundamental manifold on which the gravity theory is formulated and think of the~dS$_4$ spacetime only as a smooth limit of it. When such back-reaction is taken into account, we shall refer to the static observer as a \emph{massive observer}, and we shall think of the $q\to1$ limit as its massless probe limit in which one recovers the original, non-singular dS$_4$ geometry. 

The aim of this note is to show that massive observers in~dS$_4$ admits an intrinsic description in terms of a two-dimensional conformal field theory. We shall argue that the introduction of an orbifold parameter $q>1$ permits to build up a reduced two-dimensional action functional with support on the pair of codimension two minimal surfaces that define the set of fixed points of the $\mathbb Z_q$ action.  
Each of these minimal and tensionful surfaces have the topology of a 2-sphere in the Euclidean geometry and they can be formally thought of as the ``worldvolume" of a massive observer, whose massless limit is equivalent to the tensionless limit~$q\to1$. 
As we shall argue, the resulting effective two-dimensional Euclidean action can be identified with a Liouville theory on a 2-sphere, in which the Liouville field acquires $q$-dependent vacuum expectation value. 

The correspondence between the effective action of a massive observer and the Liouville theory action links the gravitational parameters, namely the dS$_4$ radius $\ell$ and the four-dimensional Newton's constant $G_4$, with the Liouville coupling constant $\gamma^2\sim\hbar$. This relation results in a semiclassical central charge given by
\begin{equation}\label{cintro}
c_q =\Big(1-\frac{1}{q}\Big) \frac{3\ell^2}{G_4}\,.
\end{equation}
This $q$-dependent central charge arguably encodes degrees of freedom associated to a massive observer which are not present in the massless limit~$q\to1$. 
Consequently and by means of the thermal Cardy formula, the central charge~\eqref{cintro}  predicts a Cardy entropy that equals a modular free energy whose corresponding modular entropy correctly reproduces the Gibbons--Hawking area law.

We conclude by observing that in the $q\to\infty$ limit of the quotient dS$_4/\mathbb Z_4$, the four-dimensional geometry reduces to the global geometry of dS$_3$, where the two minimal surfaces of the former can be mapped---via a double Wick rotation---to the two conformal boundaries of the latter. In this limit, the two Liouville theories on the bulk minimal surfaces become boundary theories (one for each boundary) at zero temperature. Moreover, upon taking the $q\to\infty$ limit, the two dS$_3$ boundaries inherit a central charge from the Liouville theory on the corresponding minimal surface in one higher dimension. As we shall see, the total central charge of the two boundaries reproduces exactly the central charge derived in the context of the dS$_3$/CFT$_2$ correspondence~\cite{Park:1998qk, Banados:1998tb, Strominger:2001pn,Nojiri:2001mf,Klemm:2001ea, Cacciatori:2001un, Klemm:2002ir}.

%Then, the Cardy entropy is in perfect match with its semiclassical counterpart \cite{Maldacena:1998ih, Park:1998qk, Banados:1998tb}.
%
%In $AdS_3$, the Chern-Simons formulation consist in an $SL(2,\mathbb R)\times SL(2,\mathbb R)$ gauge group, with a conformal boundary at infinity. Using the Drinfeld-Sokolov Hamiltonian reduction of the WZNW boundary theory, Coussaert, Henneaux and van Driel have shown that the asymptotic dynamics of three-dimensional gravity can be effectively described by the Liouville theory in the semiclassical limit \cite{0264-9381-12-12-012}. In this formulation, the WZNW level is related with the Liouville coupling, leading to the Brown-Henneaux asymptotic central charge of $AdS_3$ \cite{Brown:1986nw}. 

%The same formalism can be used for dS$_3$ by considering an conformal spacelike boundary. In this approach, the dS$_3$ entropy is matched with the Liouville momenta \cite{Klemm:2001ea,1126-6708-2002-04-030} (with no statistical interpretation) and the central charge of the Euclidean Virasoro algebra recovers the central charge obtained in the dS/CFT context \cite{Witten:2001kn, Strominger:2001pn, Maldacena:2002vr}, by using the transformation of the quasi-local stress tensor of dS$_3$ \cite{Balasubramanian:2002aa}. 

%~~~~~~~~~~~~~~~~~~~~~~~~~~
\section{Static observers in dS$_4$}
%~~~~~~~~~~~~~~~~~~~~~~~~~~
In four dimensions, de Sitter spacetime (dS$_4$) can be viewed as a four-dimensional timelike hypersurface embedded in five-dimensional Minkowski space $\mathcal M^{1,4}$. Taking the embedding coordinates to be $X^\mu\in\mathcal M^{1,4}$, $\mu=0,...,4$, and considering the Minkowski metric
\begin{equation}
ds_{\mathcal M^{1,4}}^2 = -(dX^0)^2 + \sum_{1\leq i \leq 4} (dX^i)^2 ~,
\end{equation}
the dS$_4$ hypersurface is defined by 
\begin{equation}
\label{hyperboloid}
X_\mu X^\mu =\ell^2~,
\end{equation}
where $\ell^2$ is the dS$_4$ radius. The hyperboloid \eqref{hyperboloid} has the topology of $\mathbb R \times S^{3}$ with manifest $O(4,1)$ symmetries.
%Now, we can define a global set of coordinates for the hyperbola by just defining
%\begin{equation}
%\label{global}
%X_0=\ell \sinh (T/\ell)~, \quad 
%X_i= \ell \cosh(T/\ell) z_i ~,~~  1\leq i \leq 4~,
%\end{equation}
%%
%where $z_i$ coordinatize the unit $3$-sphere, such that $\vec{z}^{\,2} = 1$.
%%
%Then, the resulting metric for the hyperbola hat satisfy the hypersurface equation \eqref{hyperboloid} is
%%
%\begin{equation}\label{globalmetric}
%ds^2 = -dT^2 + \ell^2 \cosh^2(T/\ell)\,d\Omega^2_{3}~,
%\end{equation}
%%
%where $-\infty < T < \infty$ and $0\leq\Theta\leq\pi$. This are called global coordinates, because they cover the whole hyperbola. 

%~~~~~~~~~~~~~~~~~~~~~~~
\subsection{Massless probe observers}
\label{Subsec:Probe}
%~~~~~~~~~~~~~~~~~~~~~~~
The standard description of a static observer in dS$_4$ is obtained by parametrizing the embedding coordinates as
\begin{equation}\label{static}
X^0= \sqrt{\ell^2- \hat r^2} \sinh(\hat t/\ell)~, \quad
X^1=\sqrt{\ell^2- \hat r^2} \cosh(\hat t/\ell)~, \quad
X^i = \hat r \hat y_i ~,~~  2\leq i \leq 4 ~,
\end{equation}
where the $\hat y_i$ denote the coordinates of the unit 2-sphere. The resulting line element
\begin{equation}
\label{staticmetric}
ds^2= - \left(1-\frac{\hat r^2}{\ell^2}\right) d\hat t^2 
          + \frac{d\hat r^2}{1 -  \frac{\hat r^2}{\ell^2}}    
          + \hat r^2 d\Omega_{2}^2  ~,
\end{equation}
where the radial coordinate runs from $0\leq \hat r<\ell$ and $d\Omega_{2}^2$ is the metric on the unit 2-sphere. 

The time-independent metric~\eqref{staticmetric} describes the worldline of a \emph{single} static observer located at the origin $\hat r=0$. The observer is causally connected with only part of the full spacetime. Such region is dubbed the Rindler wedge (or static patch) of the observer, and its boundary defines an observer-dependent cosmological horizon $\mathcal H$. This has the fix time topology of a 2-sphere and is located at $\hat r=\ell$.

In the Euclidean vacuum, a static observer detects a temperature and a corresponding Gibbons--Hawking entropy~\cite{Gibbons:1977mu} given by
\begin{equation}\label{GH}
T_{\rm dS} = \frac{1}{2\pi\ell}~,\quad \mathcal S_{\rm dS} = \frac{\pi\ell^2}{G_4}~.
\end{equation}

%~~~~~~~~~~~~~~~~~~~~~~~~~~~~~~~~~~~~~~
\subsection{Massive observers and antipodal defects}
\label{Subsec:Massive}
%~~~~~~~~~~~~~~~~~~~~~~~~~~~~~~~~~~~~~~
The above characterization of a static observer in dS spacetime considers the observer as a massless probe object. Here, instead, we treat an observer as a massive object which modify the local geometry of the spacetime; we propose to model the back-reaction of such massive observer by means of a singular~$\mathbb Z_q$ quotient. This construction, which we shall now briefly review, has been spelled out in full detail in~\cite{Arias:2019pzy}.

To begin with, we note that the constraint~\eqref{hyperboloid} can be alternatively solved by parameterizing the embedding coordinates as
\begin{align}
\label{embedding}
X_0 = &\sqrt{\ell^2 - \xi^2}\,\cos\theta\,\sinh(t/\ell)~,\quad
X_1 =\sqrt{\ell^2 - \xi^2}\,\cos\theta\,\cosh(t/\ell)~,  \\ \nonumber
X_2 &= \xi\,\cos\theta ~,\quad
X_3 =\ell \sin\theta \cos\phi~,\quad
X_4 = \ell \sin\theta \sin\phi ~,
\end{align}
where 
\begin{equation}\label{domain}
-\infty<t<\infty~,\quad 
-\ell<\xi<\ell~,\quad 
0\leq \theta\leq\pi~,\quad  
0\leq\phi<2\pi~.
\end{equation}
The resulting dS$_4$ line element, that we shall simply denote by $g_4$, is
\begin{equation}
\label{metric}
g_4=\ell^2 ( d\theta^2 + \sin^2\theta \, d\phi^2 )
+\cos^2\theta \left[- \left(  1 -  \frac{\xi^2}{\ell^2}\right) dt^2 
+ \frac{d\xi ^2}{1 -  \frac{\xi^2}{\ell^2}} \, \right]\,.
\end{equation}
The metric~\eqref{metric} has the warped product form $S^2\times_w {\rm dS}^{\pm}_2$, where the 2-sphere has radius $\ell$ and  ${\rm dS}^{\pm}_2$ denotes the radially extended dS$_2$ space, with the extended radial coordinate $\xi\in(-\ell,\ell)$, as indicated in~\eqref{domain}. This geometry describes the worldline of two antipodal static observers  
\begin{equation}\label{obs}
\mathcal O_N:= (\theta=0, \xi=0) \in \mathfrak R_N~, \qquad
\mathcal O_S:= (\theta=\pi, \xi=0) \in \mathfrak R_S~,
\end{equation}
which are causally disconnected (as any light ray can not be sent from one observer into the other). The foliation~\eqref{metric} covers the union $\mathfrak R_N \cup \mathfrak R_S$ of both northern and southern Rindler wedges, as depicted in Figure~\hyperlink{Fig:1}1.
%
%\vspace{0.5cm}
\begin{center}
\begin{tikzpicture}[scale=0.7]
\fill[fill=yellow!10] (0,0)--(2,2)--(0,4);
\fill[fill=blue!10] (4,0)--(2,2)--(4,4);
%Iplus
\node [darkred] at (0,2) {\textbullet};
\node [darkred] at (4,2) {\textbullet};
\draw [thick] (0,4) --node[above] {$\mathbf{\mathcal I^+}$}(4,4);
\draw [thick] (0,0) --node[below] {$\mathbf{\mathcal I^-}$}(4,0);                        
\draw [thick] (4,0)--(4,4); 
\node at (4.6,2) {$\mathcal O_S$};
\draw [thick] (0,0) --(0,4);
\node at (-0.6,2) {$\mathcal O_N$};
\node at (1,2) {$\mathfrak R_N$};
\node at (3,2) {$\mathfrak R_S$};
\draw [thick, dashed] (0,0) --(4,4); 
\draw [thick, dashed] (0,4)--(4,0);
%Caption~~~~~~~~~~~~~~~~~~~~
\node[text width=16cm, text justified] at (2,-4){
\small{\hypertarget{Fig:1}\bf Fig.~1}: 
\sffamily{
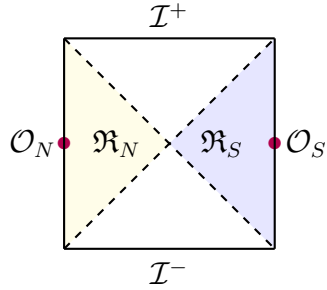
Penrose diagram of dS$_4$, with coordinates $(\tau, \Theta)$.
In the conformal time $\tau\in [-\pi/2,+\pi/2]$, the future and past infinities $\mathcal I^\pm$ are located at $\tau=\pm\pi/2$.
The global polar coordinates $\Theta\in[0,\pi]$ defines the north and south poles by
the points $\Theta=0, \pi$, respectively.
The metric~\eqref{metric} covers the two Rindler wedges of the dS$_4$ interior: $\mathfrak R_N=\{0\leq\theta<\pi/2\}$ and $\mathfrak R_S=\{\pi/2<\theta\leq\pi\}$.
The location of the two antipodal observers $\mathcal O_N$ and $\mathcal O_S$ defined in~\eqref{obs} coincides with the global north and south poles $\Theta=\theta=0$ and $\Theta=\theta=\pi$, respectively.
}};
\end{tikzpicture}
\end{center}

In order to incorporate the observers back-reaction, one next deforms the $S^2$ sector in~\eqref{metric} by performing a $S^2/\mathbb Z_q$ orbifold. This is done via the discrete identification  $\phi\sim\phi+\frac{2\pi}{q}$, with an orbifold parameter $q>1$.  
The four-dimensional orbifold $\widehat{\rm dS}_4:={\rm dS}_4/\mathbb Z_q$ is then endowed with the metric
\begin{equation}\label{g4}
\widehat g_4 = \ell^2 g_{\text{spindle}} + w^2 g_2^\pm~,
\end{equation}
where the warp factor $w = \cos\theta$ satisfy the holonomy conditions $w\rvert_{0,\pi} = 1$ and $w'\rvert_{0,\pi} = 0$, and 
\begin{equation}\label{split}
g_{\rm spindle}= d\theta^2 + \frac{\sin^2 \theta}{q^2} \,d\phi^2~,\qquad
g_2^\pm = -\left(1-\frac{\xi^2}{\ell^2}\right) dt^2 
+ \frac{d\xi ^2}{1 -  \frac{\xi^2}{\ell^2}}~.
\end{equation}

The azimuthal identification deforms the $S^2$ geometry into that of a Thurston's \emph{spindle}~\cite{thurston2014three}. The latter geometry has two antipodal conical singularities at the points $\theta=0, \pi$, which are precisely the locations of the two static observers~\eqref{obs}.  We interpret these singularities as the response of the background geometry to the presence of a massive observer, with a mass proportional to $(q-1)$.

The set of fixed points under the $\mathbb Z_q$ action 
\begin{equation}\label{fixed}
\mathcal F= \Sigma_N\cup\Sigma_S~, \qquad 
\Sigma_N := \widehat{\rm dS}_4\big|_{\theta=0}~,\qquad  
\Sigma_S := \widehat{\rm dS}_4\big|_{\theta=\pi}~,  
\end{equation}
defines two antipodal, codimension two surfaces $\Sigma_N$ and $\Sigma_S$, both endowed with the induced metric 
\begin{equation}\label{hSigma}
h=\widehat g_4 \big|_{\theta=0, \pi} = g_{2}^\pm\, .
\end{equation}
In what follows, we shall refer to the submanifolds $(\Sigma_N, h)$ and $(\Sigma_S, h)$ as \emph{defects}.

In terms of the gravity action and in order to have a well defined variational principle, the two conical singularities are resolved by adding to the Einstein--Hilbert action a pair of Nambu--Goto terms with support on $\Sigma_N$ and $\Sigma_S$~\cite{Fursaev:1995ef}

\begin{equation}\label{Itotal}
I[\widehat{\rm dS}_4] =\frac{1}{16\pi G_4} 
{\displaystyle\int\limits_{\widehat{\rm dS}_4\setminus(\Sigma_N\cup\Sigma_S)}}
d^4 x \sqrt{-g}\Big( R- \frac{6}{\ell^2}\Big) 
-\mathcal T_q\int_{\Sigma_N} d^2 y\sqrt{-h}
-\mathcal T_q\int_{\Sigma_S} d^2 y\sqrt{-h}~.
\end{equation}
In the above, the support of the first integral excludes the location of the defects $\Sigma_N$ and $\Sigma_S$. The two Nambu--Goto terms are coupled through the tension
\begin{equation}\label{tension}
\mathcal T_q = \frac{1}{4 G_4} \Big(1-\frac{1}{q}\,\Big)\,,
\end{equation} 
where the limit $q\to1$ corresponds to the tensionless limit in which one recovers the usual Einstein--Hilbert action on the smooth dS$_4$ geometry.

Hence, by construction, $\Sigma_N$ and $\Sigma_S$ are codimension two minimal surfaces with an induced stress-energy tensor given by
\begin{equation}\label{Tij}
T_{ij}= \mathcal{T}_q \,h_{ij} \,.
\end{equation}
The localized stress energy tensor~\eqref{Tij} is a strong sign of the existence of an underlying field theory defined on the two minimal surfaces. As we shall next argue, this theory corresponds to an Eucliedan Liouville theory on a 2-sphere. 
%~~~~~~~~~~~~~~~~~~~~~~~~~~~~~~~~~~~~~~~~~~~
\section{Liouville theory description of a massive observer}
\label{Sec:3}
%~~~~~~~~~~~~~~~~~~~~~~~~~~~~~~~~~~~~~~~~~~~
In this section, we construct an effective two-dimensional action with support on the codimension two minimal surfaces $\Sigma_N$ and $\Sigma_S$. 
These surfaces are the set of fixed points of the $\mathbb Z_q$ action. Each of them contain the worldline of one of the massive observers $\mathcal O_N$ and $\mathcal O_S$, and they both have the topology of a 2-sphere in the Euclidean geometry, \emph{viz}.
\begin{equation}\label{NSS2}
\Sigma^E_N \cong \Sigma^E_S\cong S^2\,,
\end{equation}
with induced metric $d\Omega_2^2$ (which corresponds to the analytic continuation of~\eqref{hSigma}). In the above, the label ``$E$" denotes Euclidean geometry. Hereafter, we shall drop this label when is clear from context. 

%~~~~~~~~~~~~~~~~~~~~~~~~~~~~~~~~~
\subsection{Effective two-dimensional action}
%~~~~~~~~~~~~~~~~~~~~~~~~~~~~~~~~~
To begin with, we recall that the total Euclidean gravity action~\eqref{Itotal} on the conically singular manifold $\widehat{\rm dS}_4:={\rm dS}_4/\mathbb Z_q$ consists of a bulk piece plus a pair of two-dimensional Nambu-Goto terms
\begin{equation}\label{IEtotal}
I^E_{\rm total}[\widehat{\rm dS}_4]
= I_{\rm bulk}[\widehat{\rm dS}_4]
+I_{\rm NG}[\Sigma_N] +I_{\rm NG}[\Sigma_S]~, 
\end{equation}
where the Euclidean integrals
\begin{align}
I_{\rm bulk}[\widehat{\rm dS}_4]&:=-\frac{1}{16\pi G_4} 
{\displaystyle\int\limits_{\widehat{dS}_4\setminus(\Sigma_N\cup\Sigma_S)}}
d^4 x \sqrt{g}\Big( R- \frac{6}{\ell^2}\Big)~, \\\nonumber
I_{\rm NG}[\Sigma_N]:= \mathcal T_q &\int_{\Sigma_N} d^2 y\sqrt{h}~, \qquad
I_{\rm NG}[\Sigma_S]:= \mathcal T_q \int_{\Sigma_S} d^2 y\sqrt{h}~.
\end{align}

Although the support of the bulk integral above excludes the location of the defects, we can define a ``free energy inflow" from the bulk to $\Sigma_N$ and $\Sigma_S$ by dimensional reducing $I_{\rm bulk}$ down to two dimensions 
\begin{equation}\label{dimred}
I_{\rm bulk}[\widehat{\rm dS}_4] 
\stackrel{\substack{\rm dim\,red\vspace{-1mm}\\{}}}{\longrightarrow}
I_{2d}[\Sigma_N\big] 
+ I_{2d}[\Sigma_S\big]~,
\end{equation}
as to define an effective action on each of the defects, which comprises the inflow~\eqref{dimred} and the corresponding Nambu--Goto term, \emph{viz}.
\begin{equation}\label{effNS}
I_{\rm eff}[\Sigma_N] 
= I_{2d}[\Sigma_N] 
+ I_{\rm NG}[\Sigma_N] ~,\quad
I_{\rm eff}[\Sigma_S] 
= I_{2d}[\Sigma_S] 
+ I_{\rm NG}[\Sigma_S]~, 
\end{equation}
and such that the total on-shell action~\eqref{IEtotal}
\begin{equation}\label{totaleff}
I^{E}_{\rm total}[\widehat{\rm dS}_4]
\approx I_{\rm eff}[\Sigma_N] 
+ I_{\rm eff}[\Sigma_S] ~.
\end{equation}
(From here and in what follows, we shall use the notation ``$\approx$" to indicate on-shell equalities.)

The reduced Euclidean action $I_{2d}$ in~\eqref{dimred} can be computed using the line elements~\eqref{split} and integrating out the spindle coordinates $(\theta, \phi)$. This gives
\begin{equation}
\label{4d2d}
I_{\rm bulk}[\widehat{\rm dS}_4]
\approx  
-\frac{\ell^2}{4qG_4} \int d^2y\sqrt h\,\mathcal R ~,
\end{equation}
where the integral is over the two-dimensional submanifold coordinatized by $y=(\tau, \xi)$ (with $\tau$ denoting the Euclidean time), and $\mathcal R=\mathcal R[h]$ is the intrinsic two-dimensional scalar of curvature built up from the induced metric on the defects~\eqref{hSigma}. This reduction holds upon imposing Einstein's equations 
$\ell^2 R_{\theta\theta}=3g_{\theta\theta}$ 
(and likewise the $\phi\phi$-equation) and by making use of the codimension two identity 
$R_{ij}=3\cos^2\theta\, \mathcal R_{ij}$. 

Due to the antipodal symmetry relating $\Sigma_N$ and $\Sigma_S$~\cite{Arias:2019pzy}, we further assign to each of the defects half of the total inflow~\eqref{4d2d}
\begin{equation}
I_{2d}[\Sigma_N]:= 
-\frac{\ell^2}{8 q G_4} 
\int_{\Sigma_N} d^2y\sqrt h\,\mathcal R~,\quad 
I_{2d}[\Sigma_S]:= 
-\frac{\ell^2}{8 q G_4} 
\int_{\Sigma_S} d^2y\sqrt h\,\mathcal R~,
\end{equation}
so that the effective action~\eqref{effNS} on the northern defect is given by
\begin{equation}\label{IeffN}
I_{\rm eff}[\Sigma_N] \approx 
- \frac{\ell^2}{8qG_4} 
\int_{\Sigma_N} d^2y\sqrt h\,\mathcal R
+\frac{1}{4 G_4} \Big(1-\frac{1}{q}\Big)
\int_{\Sigma_N} d^2y\sqrt h ~,
\end{equation}
\emph{idem} for $I_{\rm eff}[\Sigma_S]$.

\subsection{On-shell correspondence with Liouville theory}
We now observe that the structure of the reduced effective action~\eqref{IeffN}  closely resembles the Liouville theory action~\cite{Polyakov:1981rd}:
\begin{equation}\label{IL}
I_{\rm L}[g, \Phi; \gamma]= -\frac{1}{2} \int_{M_2} d^2y \sqrt g 
\Big(g^{ij}\partial_i\Phi \partial_j\Phi + Q\mathcal R \Phi 
+4\pi \mu e^{2\gamma\Phi}\Big)~.
 \end{equation}
In the above, $(M_2, g)$ is a two-dimensional Euclidean manifold 
and $\gamma^2\sim\hbar$ is the only coupling constant 
of the theory; its strength dictates the classical and quantum regimes 
and further defines the background charge to be $Q=\gamma^{-1} + \gamma$, 
as required for conformal invariance (for a brief review of Liouville 
theory see~Appendix~\ref{App:A}). 
It is important to point out that the action~\eqref{IL} differs from the one given in~\eqref{Qaction} by an overall factor of $-2\pi$; such a normalization is 
needed in order to uniformize the definition of the stress-energy tensor\footnote{We recall that in our conventions the definition of the gravitational stress energy tensor~\eqref{Tij} differs from the standard convention used in the CFT context:
$$
T^{ij}_{\rm grav} = \frac{2}{\sqrt{h}} \frac{\delta I}{\delta h_{ij}}~, \qquad
T^{ij}_{\rm CFT} = \frac{-4\pi}{\sqrt{h}} \frac{\delta I}{\delta h_{ij}}~.
$$
The overall factor of $-2\pi$ propagates when computing the operator product expansion $\langle TT\rangle$, which in turns produces a relative factor in the central charge. Comparison of the effective gravitational action~\eqref{IeffN} with the Liouville action~\eqref{Qaction} thus requires the normalization implemented in~\eqref{IL}.} while comparing~\eqref{IeffN} and~\eqref{IL}.

Indeed, the reduced effective action~\eqref{IeffN} corresponds precisely 
to the Liouville action~\eqref{IL} on the 2-sphere $(M_2, g)=(\Sigma_N, h)$, upon giving to the Liouville field a fix expectation value~$\langle \Phi\rangle= \Phi_0$. 
That is
\begin{equation}\label{duality}
I_{\rm eff}[\Sigma_N]
\approx 
I_{\rm L}\big|_{\langle \Phi\rangle= \Phi_0}~,
\end{equation}
and similarly for $I_{\rm eff}[\Sigma_S]$. This on-shell relation permits to establish the existence of an effective field theory on each of the minimal surfaces, $\Sigma_N$ and $\Sigma_S$, given by a broken phase of Liouville theory.  As we shall now see, the effective field theoretic description encode a number of compatibility conditions that in turn yield a $q$-dependent central charge which we propose encode the degrees of freedom associated to a massive observer\footnote{A similar idea has been previously discussed in \cite{Solodukhin:1998tc} where the dimensional reduction of Einstein gravity to two-dimensional Liouville theory is proposed to describe the underlying degrees of freedom of black hole horizons.}.

The on-shell correspondence~\eqref{duality} holds provided 
\begin{equation}\label{matching1}
\frac{\ell^2}{8qG_4}=\frac{Q\Phi_0}{2}~,\quad
\frac{1}{4G_4}\Big(1-\frac{1}{q}\Big)
=-2\pi\mu e^{2\gamma\Phi_0}~,
\end{equation}
as follows from matching the terms of the same order in derivatives of the metric in~\eqref{IeffN} and~\eqref{IL}.
In addition, the expectation value $\Phi_0$ must satisfy the Liouville equation of motion for a constant field, which is given by 
\begin{equation}\label{matching2}
Q\mathcal R + 8\pi \gamma \mu e^{2\gamma\Phi_0} 
= \frac{2Q}{\ell^2} + 8\pi \gamma \mu e^{2\gamma\Phi_0}=0~,
\end{equation}
where the first equality made use of the constant 
positive curvature $\mathcal R=2\ell^{-2}$ of $\Sigma_N$. 

Compatibility of the equations~\eqref{matching1} and~\eqref{matching2} yields 
\begin{equation}\label{Pg}
\Phi_0=\frac{1}{2\gamma(q-1)}~, \qquad 
\mu=\frac{1}{8\pi G_4}\bigg( \frac{1-q}{q}\bigg) \exp\bigg(\frac{1}{1-q}\bigg)~,
\end{equation}
and
\begin{equation}\label{gq}
\frac{Q}{\gamma} = 
\Big(1-\frac{1}{q}\Big)\frac{\ell^2}{2 G_4}~.
\end{equation}

Observe that the bound $q>1$ for the orbifold parameter can be understood as a consistency condition: One the one hand, from~\eqref{matching2} it follows 
that positivity of $\mathcal R[h]=2\ell^{-2}>0$ is only possible if 
$\mu<0$, which according to~\eqref{Pg} requires $q$ to be greater than 
one. On the other hand and remembering that $Q=\gamma+\gamma^{-1}$, the  bound $q>1$ ensures the reality of the couplings~$\gamma$ and~$\ell$ in~\eqref{gq}.
 
%{\color{blue} In the quantum local analysis of the Liouville theory \cite{DHoker:1982aa, Seiberg:1990eb}, the normalizable modes if first studied by taking the limit of $\phi\sim -\infty$ which in our case corresponds to the tensionless limit.}

In what follows, we shall see that the semiclassical limit of~\eqref{gq} provides a nontrivial link between the Liouville coupling constant $\gamma$, in terms of which the central charge of the theory is defined, and the gravitational coupling $\ell^2/G_4$ which in turns defines (up to a factor of $\pi$ in dimension four) the entropy of the dS$_4$ space. 
%~~~~~~~~~~~~~~~~~~~~~~~~~~~~~~~~~~
\subsection{Central charge and Cardy formula}
%~~~~~~~~~~~~~~~~~~~~~~~~~~~~~~~~~~
In the semiclassical regime $\gamma^2\ll1$, 
where thus $Q\sim\gamma^{-1}$, there exists 
a $\mathcal O(1/\gamma^2)$ contribution to the 
Liouville central charge~\cite{Jackiw:2005su}
\begin{equation}
c = 1+6Q^2\approx \frac{6}{\gamma^2}~,
\end{equation}
whose value can be computed in terms of the gravity
couplings and the orbifold parameter $q$. Indeed, from the semiclassical limit of~\eqref{gq}, we straightforwardly find 
\begin{equation}\label{cq}
c_q= \Big(1-\frac{1}{q}\Big)\frac{3\ell^2}{G_4}~.
\end{equation}
This value of the central charge is consistent with the classical conformal anomaly equation $12\langle T \rangle= c\mathcal{R}$, where $T=h^{ij}T_{ij}$ is the trace of stress-energy tensor~\eqref{Tij} and $\mathcal R=2\ell^{-2}$ is the curvature of the corresponding defect.
Also, we note that since $q>1$, then the central charge~$c_q > 0$, which indicates unitarity of the theory.

Is it worth to notice that the central charge~\eqref{cq} is $q$-dependent and vanishes in the tensionless limit $q\to1$. This is similar to what occurs in the AdS$_3$/CFT$_2$ context, where the bulk orbifold ${\rm AdS}_3/\mathbb Z_q$ induces a $q$-dependence of the central charge of the boundary theory \cite{Balasubramanian:1999zv, Martinec:2002xq, Balasubramanian:2003kq, deBoer:2010ac, Chen:2018vkw}, with the Brown--Henneaux central charge being recovered by rescaling the Newton's constant as $G\sim qG$.

Having obtained the central charge~\eqref{cq} and by virtue of
the thermal Cardy formula in the canonical ensemble~\cite{Cardy:1986ie, Bloete:1986qm}
\begin{equation}\label{CardyE1}
{\mathcal S}_q^{\,\rm Cardy} 
= \frac{\pi^2}{3} c_{q, L}\, T_L 
+ \frac{\pi^2}{3} c_{q, R}\, T_R~,
\end{equation}
a $q$-dependent entropy can be computed (as usual, 
$L$ and $R$ label left and right-movers central charge 
and temperature). 
Indeed, based on the arguments of~\cite{Hartman:2014oaa}, the Cardy formula holds in the extended range of large central charge and large gap in operator dimension above zero. Accordingly, in our case, since
\begin{equation}
c_q\sim\frac{\ell^2}{G_4} \gg 1~,\quad
\Delta_0 \sim c_q \gg1~,
\end{equation}  
where $\Delta_0$ is the (semiclassical) conformal dimension of the bound state (see~\eqref{cD}), the Cardy formula~\eqref{CardyE1} applies.

For a non-chiral Liouville theory, we have
\begin{equation}\label{cT}
c_{q, L}=c_{q, R}=c_q ~, \qquad
T_L=T_R=\frac{1}{2\pi}~,
\end{equation}
where $c_q$ is given in~\eqref{cq} and $T_L$ and $T_R$ correspond to the temperature of the generalized Hartle--Hawking vacuum of dS space. This is known to be equivalent to a thermal state $\rho=e^{-2\pi H_{\mathfrak R}}$ defined by the Rindler Hamiltonian $H_{\mathfrak R}$~\cite{Laflamme:1988wg,Barvinsky:1994jca}(see also~\cite{Jacobson:1994fp}). 

Hence, using~\eqref{cq} and~\eqref{cT} in the Cardy formula~\eqref{CardyE1}, we find the $q$-dependent Cardy entropy
\begin{equation}\label{CardyE}
\mathcal S_q^{\,\rm Cardy}
= \Big(1-\frac{1}{q}\Big)\frac{\pi\ell^2}{G_4} ~.
%=\Big(1-\frac{1}{q}\Big)\frac{\mathcal A_{\mathcal H}}{4G_4} ~.
\end{equation}
Note that minus the derivative of the above entropy with respect to $1/q$ gives the Gibbons--Hawking entropy~\eqref{GH}. Based on this simple observation, we shall next reinterpret the Cardy entropy~\eqref{CardyE} as \emph{modular free energy}. 

\subsection{Modular free energy and Gibbons--Hawking entropy}
{\color{red} }
The Cardy entropy~\eqref{CardyE} can be understood as the modular free energy $F_q$ whose derivative with respect to the dimesionless temperature\footnote{Note that, by identifying $T=q^{-1}$, the orbifold parameter $q$ induces a Boltzmann factor $\exp(-qH)$ (with $H$ denoting the modular Hamiltonian~\eqref{Hmod}) which makes the Liouville theory thermal and thus amounts to using the thermal Cardy formula~\eqref{CardyE1}.} $T=q^{-1}$~\cite{Baez:2011, Rangamani:2016dms} yields the Gibbons--Hawking area law. To this end, we define the modular Hamiltonian
\begin{equation}\label{Hmod}
H := -\log\rho~, \qquad \rho^{\,q}=e^{-qH}~.
\end{equation}
Thus, we can write the modular partition function as
\begin{equation}
\mathcal Z={\rm tr}\,\rho^{\,q}={\rm tr}\,e^{-qH}~,
\end{equation}
in terms of which the modular free energy is given by
\begin{equation}\label{Fq}
F_q=-\frac{1}{q} \log \mathcal Z= -\frac{1}{q}\log {\rm tr}\rho^{\,q}~.
\end{equation}

Next, we compute the modular free energy~\eqref{Fq} on the 4-sphere defined by
\begin{equation}\label{S4}
S^4:= (\mathcal R^E_S\cup\mathcal R^E_N)/\Pi \,.
\end{equation}
Here, $\mathcal R^E_S$ and $\mathcal R^E_N$ denote the analytic continuation of the southern and northern Rindler wedges (both given by 4-spheres), and $\Pi: S^4\rightarrow S^4$ is the antipodal map that sends every point in the southern Rindler wedge to the corresponding antipodal point in the northern Rindler wedge~\cite{Arias:2019pzy}. It is important to point out that the 4-sphere~\eqref{S4} is equivalent to the analytic continuation of a single Rindler wedge and also equivalent to the Euclidean continuation of global ${\rm dS}_4$ spacetime. Moreover, it naturally admits a $\mathbb Z_q$ action (given by azimutal identifications), with a $q$-fold branched cover that we denote by $S^4_q$. 
Using the Calabrese-Cardy formula~\cite{Calabrese:2004eu}
\begin{equation}\label{CC}
{\rm tr}\,\rho^{\,q} = \frac{\mathcal Z[S^4_q]}{(\mathcal Z[S^4])^q}\,,
\end{equation}
it follows that
\begin{equation}\label{Fq1}
F_q[S^4] = -\frac1q \log \mathcal Z[S^4_q] + \log \mathcal Z[S^4] 
\approx I^E[S^4/\mathbb Z_q]- I^E[S^4] 
= 2\,\Big(1-\frac{1}{q}\Big)\frac{\pi\ell^2}{G_4}\,.
\end{equation}
In the above, we have used the semiclassical approximation $\mathcal Z[S^4]\approx \exp(-I^E[S^4])$ and the locality of the gravity action to write $I^E[S^4_q]=qI^E[S^4/\mathbb Z_q]$. The value of the latter is given by the on-shell value of~\eqref{Itotal} (properly Euclideanized) restricted to a single Rindler wedge (which we recall is given by a 4-sphere in the Euclidean geometry), \emph{viz.}
\begin{equation}
I^E[S^4/\mathbb Z_q] \approx \Big(1-\frac{2}{q}\Big)\frac{\pi\ell^2}{G_4}\,.
\end{equation}

The value of the modular free energy~\eqref{Fq1} comprises the contribution form both, northern and southern defects. For a single defect (say the southern one), we thus have
\begin{equation}\label{Fq2}
F_q^{\Sigma_S} = \frac{1}{2} F_q[S^4] = \Big(1-\frac{1}{q}\Big)\frac{\pi\ell^2}{G_4}\,.
\end{equation}
which corresponds exactly to the value of the Cardy entropy~\eqref{CardyE}.

Finally, we can compute the modular entropy 
\begin{equation}\label{thermalS}
\widetilde{\mathcal S}_q = -\frac{\partial F_q}{\partial T} 
= (1 - q\partial_q)\log\mathcal Z\,,
\end{equation}
which gives
\begin{equation}\label{Smod}
\widetilde{\mathcal S}_q
=q^2\frac{\partial}{\partial q}  \Big(1-\frac{1}{q}\Big)\frac{\pi\ell^2}{G_4} 
=\frac{\pi\ell^2}{G_4}=\mathcal S_{\rm dS}\,.
\end{equation}
This is precisely the Gibbons--Hawking entropy~\eqref{GH}. Observe that although~\eqref{Smod} has its origin in the modular free energy~\eqref{Fq}, its value is independent of the modular parameter $q$ and hence this remains fix in the tensionless 
limit $q\to1$, in which one recovers the standard description of the dS$_4$ spacetime.
\section{The large $q$ limit and dS$_3$ holography}
\label{Sec:4}
%~~~~~~~~~~~~~~~~~~~~~~~~~~~~~~~~~~~~~~~
%The formulation of a gravity theory on the orbifold $\widehat{\rm dS}_4:={\rm dS}_4/\mathbb Z_q$ bring to the fore multiple sectors of the theory defined by the various limits of the orbifold parameter $q$.  In this setup, seemingly disparate geometries turns out to be given by different limits of the same parent manifold $\widehat{\rm dS}_4$, which in turns amount to treat all such geometries on an equal footing. 
%
Here, we consider the $q\to\infty$ limit of the orbifold $\widehat{\rm dS}_4:={\rm dS}_4/\mathbb Z_q$. We will first argue that this limit yields an alternative realization of the global ${\rm dS}_3$ geometry. We will further propose that the large $q$ limit provides a new mechanism to study ${\rm dS}_3/{\rm CFT}_2$ holography, whereby the dual field theory defined on the two conformal boundaries of ${\rm dS}_3$ has a higher dimensional origin, namely, it is inherited from the Euclidean Liouville theory on the two minimal surfaces~$\Sigma_S$ and~$\Sigma_N$ (embedded in four dimensions).

\subsection{3D conformal boundaries from codimension two defects in 4D}
The limit $q\to\infty$ is equivalent to the zero radius limit of the $S^2/\mathbb Z_q$ spindle, \emph{viz.} $\ell_q:=q^{-1}\ell\to0$. In this limit, the two-dimensional geometry between the northern and southern defects~$\Sigma_N$ and~$\Sigma_S$ collapses to a single transverse direction, say $z:=\ell\theta$, with $\Sigma_N$ located at $z=0$ and $\Sigma_S$ at $z=\pi\ell$. 
The situation is illustrated in Figure~\hyperlink{Fig:2}2. 
\vspace{0.7cm}
\begin{center}
\begin{tikzpicture}[scale=0.7]
%2-sphere
\draw [thick] (-4,0) circle (2cm);
\node at (-5.5,2) {\footnotesize$S^2$};
\draw[thick, dashed] (-2,0) 
arc[start angle=0,end angle=180, 
x radius=2, y radius=0.5];
\draw[thick] (-2,0) 
arc[start angle=0,end angle=-180, 
x radius=2, y radius=0.5];
%Gamma
{\color{darkred}
\draw[thick, dashed] (-2.4,1.2) 
arc[start angle=0,end angle=180, 
x radius=1.6, y radius=0.2];
\draw[thick] (-2.4,1.2) 
arc[start angle=0,end angle=-180, 
x radius=1.6, y radius=0.2];
\node at (-2.1,1.5) {\footnotesize{$\Gamma_\theta$}};
}
%Orbifold arrow
{\color{darkblue}
\node at (-0.5,0.45) {\footnotesize{$S^2/\mathbb Z_q$}};
\draw [thick, ->] (-1.3,0) -- (0.35,0);
}
%Spindle
\fill[fill=yellow!10] (2,2.3)--(1,1.7)--(6,1.7)--(7,2.3);
\fill[fill=blue!10] (2,-1.7)--(1,-2.3)--(6,-2.3)--(7,-1.7);
\draw [thick] (2,2.3) --(7,2.3);
\draw [thick] (1,1.7) --(6,1.7);
\draw [thick] (1,1.7) --(2,2.3);
\draw [thick] (7,2.3) --(6,1.7);
\coordinate (N) at (4,2);\coordinate (S) at (4,-2);
\node [darkred] at (N) {\textbullet};
\node [darkred] at (S) {\textbullet};
\draw[thick] (N)to[out=-20,in=20](S);
\draw[thick] (N)to[out=-150,in=150](S);
%\draw [thick](3.05,0) ellipse (1.05cm and 0.3cm);
\draw[thick, dashed] (5.1,0) 
arc[start angle=0,end angle=180, 
x radius=1.05, y radius=0.3];
\draw[thick] (5.1,0) 
arc[start angle=0,end angle=-180, 
x radius=1.05, y radius=0.3];
%\node at (4.4,2.9) 
%{\footnotesize{$c_q(\Sigma_N)$}$= (1-\tfrac{1}{q})\tfrac{3\ell^2}{G_4}$};
%\node at (4.4,-2.9) 
%{\footnotesize{$c_q(\Sigma_S)$}$= (1-\tfrac{1}{q})\tfrac{3\ell^2}{G_4}$};
\draw [thick] (2,-1.7) --(7,-1.7);
\draw [thick] (1,-2.3) --(6,-2.3);
\draw [thick] (1,-2.3) --(2,-1.7);
\draw [thick] (7,-1.7) --(6,-2.3);
\draw [thick, ->] (4.05,0) -- (5.05,0);
\node at (6,0) {\footnotesize{$\ell_q=\frac{\ell}{q}$}};
\node at (7,1.7) {\footnotesize $\Sigma_N$};
\node at (1,-1.7) {\footnotesize $\Sigma_S$};
%\node at (6,0) {\footnotesize{$\ell_q=\frac{\ell}{q}$}};
%Largeqarrow
{\color{darkblue}
\node at (8.4,0.35) {\footnotesize{$q\to\infty$}};
\draw [thick, ->] (7.5,0) -- (9.3,0);
}
%ZEROREADIUSLIMIT
\fill[fill=yellow!10] (10,2.3)--(9,1.7)--(14,1.7)--(15,2.3);
%\node [darkred] at (10,2) {\textbullet};
\draw [thick] (10,2.3) --(15,2.3);
\draw [thick] (9,1.7) --(14,1.7);
\draw [thick] (9,1.7) --(10,2.3);
\draw [thick] (15,2.3) --(14,1.7);
\fill[fill=blue!10] (10,-1.7)--(9,-2.3)--(14,-2.3)--(15,-1.7);
%\node [darkred] at (10,-2) {\textbullet};
\draw [thick] (10,-1.7) --(15,-1.7);
\draw [thick] (9,-2.3) --(14,-2.3);
\draw [thick] (9,-2.3) --(10,-1.7);
\draw [thick] (15,-1.7) --(14,-2.3);
\draw [thick, dashed] (12,-2) --(12,2);
\node at (14,0) {\footnotesize{dS$_3$}};
\draw [thick, ->] (12,2) -- (12,1);
\node at (11.7,1.3) {\footnotesize{$z$}};
%\node at (12, 2.9) 
%{\footnotesize{$c_\infty(\Sigma_N)$}$=\tfrac{3\ell}{4G_3}$};
%\node at (12,-2.9) 
%{\footnotesize{$c_\infty(\Sigma_S)$}$=\frac{3\ell}{4G_3}$};
%CAPTION
\node[text width=16cm, text justified] at (5,-5) 
{\small {\hypertarget{Fig:2}\bf Fig.~2}: 
\sffamily{
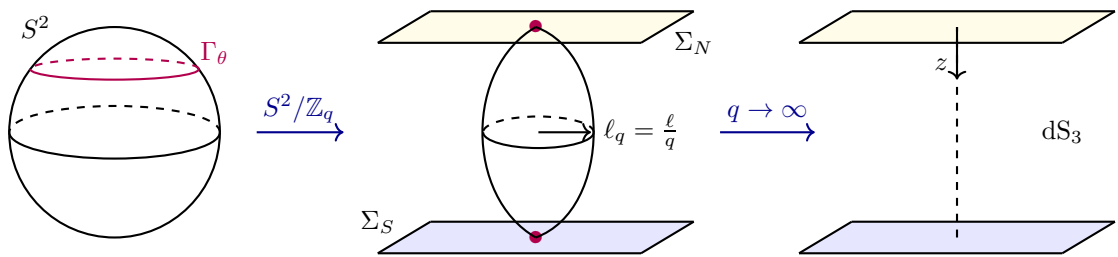
The large $q$ limit of the spindle $S^2/\mathbb Z_q$. This corresponds to the zero radius limit $\ell_q\to0$, where the two-dimensional geometry between the northern and southern defects $\Sigma_N$ and $\Sigma_S$ shrinks to a single transverse dimension. The resulting geometry is that of global ${\rm dS}_3$ spacetime.
}};
\end{tikzpicture}
\end{center}
\vspace{0.5cm}

In the above limit, the four-dimensional geometry of the manifold $(\widehat{\rm dS}_4, \widehat g_4)$ reduces to the three-dimensional geometry of global dS$_3$ spacetime with a radius equals to $\ell$. This can be seen directly from the embedding coordinates~\eqref{embedding} by first identifying $\phi\sim\phi+\frac{2\pi}{q}$ and then taking $q\to\infty$. This operation sets $X_4=0$. The remaining coordinates 
\begin{align}
\label{dS3embedding}
X_0 = \sqrt{\ell^2 - \xi^2}\,\cos\theta\,\sinh(t/\ell)~&,\quad
X_1 =\sqrt{\ell^2 - \xi^2}\,\cos\theta\,\cosh(t/\ell)~,  \\ \nonumber
X_2 = \xi\,\cos\theta ~&,\quad
X_3 =\ell \sin\theta~,
\end{align}
parametrize the embedding dS$_3\hookrightarrow \mathcal M^{1,3}$ of the dS$_3$ hyperboloid, defined by the hypersurface equation $-(X^0)^2+(X^1)^2+(X^2)^2+(X^3)^2=\ell^2$, into four-dimensional Minkowski spacetime $\mathcal M^{1,3}$.

After taking the limit, the resulting geometry is
\begin{equation}\label{g3}
g_{3} = dz^2 + \cos^2(z/\ell)\,h\,,
\end{equation}
where $h$ is the two-dimensional induced metric on the defects defined in~\eqref{split} and~\eqref{hSigma}. We futher observe that the line element~\eqref{g3} can be mapped to the global foliation of ${\rm dS}_3$.  This is done via analytical continuation of the transverse coordinate $z\in[0,\pi\ell]$ and the time $t\in(-\infty, \infty)$ (the latter being the time coordinate in $h$), that is
\begin{equation}\label{ziT}
z\rightarrow iT~,\qquad  t\rightarrow i\tau~. 
\end{equation}
As a result, the compact coordinate $z$ becomes the global time $-\infty <T<\infty$ and the induced metric $h\rightarrow d\Omega_2^2$, where $d\Omega_2^2$ denotes the metric on the unit 2-sphere: 
\begin{equation}
g_{3} = -dT^2 + \cosh^2(T/\ell) d\Omega^2_2~.
\end{equation}
Clearly, this is the global foliation of dS$_3$ spacetime. Under~\eqref{ziT},  the original codimension two defects $\Sigma_N$ and $\Sigma_S$ are  respectively sent to $T\rightarrow-\infty$ and $T\rightarrow\infty$. Hence, in the large $q$ limit, they reincarnate as the past and future infinities of dS$_3$.

\subsection{dS$_3$/CFT$_2$ central charge}
The above maneuvers show that the global dS$_3$ geometry can be thought of as the limit
\begin{align}\label{dS3limit}
(\widehat{\rm dS}_4, \widehat g_4)
&\stackrel{\substack{q\to\infty\vspace{-1mm}\\{}}}{\longrightarrow}
({\rm dS}_3, g_3) \\ \nonumber
(\Sigma_S, \Sigma_N) &\longmapsto (\mathcal I^+, \mathcal I^-)~,
\end{align}
where the minimal surfaces $\Sigma_S$ and $\Sigma_N$ are sent to the past and future infinities $\mathcal I^\pm$ of dS$_3$ (after the double analytical continuation~\eqref{ziT}). Thus, recalling from Section~\ref{Sec:3} that on $\Sigma_N$ and $\Sigma_S$ there exist an Euclidean Liouville theory, 
from the dS$_3$ perspective one should expects to have some Liouville-type theory on each of the boundaries $\mathcal I^\pm$.  
This is consistent with the known fact that the asymptotic dynamics of pure dS$_3$ gravity---when formulated as two copies 
Chern--Simons theory with gauge group SL$(2, \mathbb C)$---is described by an Euclidean Liouville theory on $\mathcal I^+\cup\mathcal I^-$~\cite{Klemm:2002ir}. Indeed, in the large $q$ limit, the Liouville theory on each the minimal surfaces reaches its zero temperature limit becoming a non-thermal theory, in agreement with the results established in~\cite{Klemm:2002ir} which, in the context of the ${\rm dS}_3/{\rm CFT}_2$ correspondence, predicts a non-thermal dual theory.

Accordingly, the total central charge of the composite boundary $\mathcal I^+\cup\mathcal I^-$  
\begin{equation}
c= c(\mathcal I^+) + c(\mathcal I^-)\,, 
\end{equation}
can be computed by means of~\eqref{dS3limit} as
\begin{equation}\label{cinfty}
c= c_\infty(\Sigma_N) + c_\infty(\Sigma_S)=\frac{6\ell^2}{G_4}\,,
\end{equation}
where $c_\infty(\Sigma_N)=c_\infty(\Sigma_S)=\frac{3\ell^2}{G_4}$ denote the Liouville central charge~\eqref{cq} in the limit $q\to\infty$. Note that the four-dimensional Newton's constant can be expressed in terms of the three-dimensional one as
\begin{equation}
G_4={\rm Vol}(S^1)\,G_3\,,
\end{equation}
where ${\rm Vol}(S^1)$ is defined as the average volumen of a meridian $\Gamma_\theta$ located at a polar angle $\theta$ (see Figure~\hyperlink{Fig:2}2). This average is given by
\begin{equation}
{\rm Vol}(S^1)= \langle \Gamma_\theta \rangle=2\pi\ell\langle\sin\theta\rangle=4\ell\,.
\end{equation}
(In the above, we have used that  $\pi\langle\sin\theta\rangle=\int_0^\pi d\theta\sin\theta=2$.) Then
\begin{equation}\label{G4G3}
G_4=4\ell G_3\,, 
\end{equation}
and therefore one finds that the total central charge~\eqref{cinfty} is
\begin{align}\label{cBH}
c= \frac{3\ell}{2G_3}\,,
\end{align}
in accordance with the result derived in the context of the dS$_3$/CFT$_2$ correspondence~\cite{Park:1998qk, Banados:1998tb, Strominger:2001pn,Nojiri:2001mf,Klemm:2001ea, Cacciatori:2001un, Klemm:2002ir}. 
Note that this result is consistent with the fact that, in the large $q$ limit, the Cardy entropy of the two defects
\begin{equation}
\mathcal S_q^{\Sigma_N\cup\Sigma_S}:=\mathcal S_q^{\,\rm Cardy}[\Sigma_N] +  \mathcal S_q^{\,\rm Cardy}[\Sigma_S] 
= 2\left( 1 - \frac1q\right)\frac{\pi\ell^2}{G_4}~,
\end{equation}
correctly reproduces the thermodynamic entropy of three-dimensional dS spacetime~\cite{Maldacena:1998ab} (upon using the dimensional reduction of the Newton constant \eqref{G4G3}), \emph{viz}.
\begin{equation}\label{SdS3}
\mathcal S_q^{\Sigma_N\cup\Sigma_S} 
\stackrel{\substack{q\to\infty\vspace{-1mm}\\{}}}{\longrightarrow}
 \mathcal S_{{\rm dS}_3}=\frac{\pi\ell}{2G_3} \,.
\end{equation}
The results~\eqref{cBH} and~\eqref{SdS3} seem to indicate that dS$_3$ holography may emerge as the large $q$ limit of the ${\rm dS}_4/\mathbb Z_q$ orbifold geometry.

%~~~~~~~~~~~~~~~~~~~~~~~~~
\section{Conclusions}\label{Sec:5}
%~~~~~~~~~~~~~~~~~~~~~~~~~
In this work, we have modeled the back-reaction of a static observer in four-dimensional de Sitter spacetime via the singular quotient~dS$_4/\mathbb Z_q$. The latter geometry exhibits two antipodal conical singularities that we interpret as being created by a pair of massive observers, $\mathcal O_S$ and $\mathcal O_N$, defined in~\eqref{obs}. The massless probe limit is defined by $q\to1$ in which one recovers the smooth dS$_4$ spacetime. 

The set of fixed points of the $\mathbb Z_q$ action defines a pair of codimension two surfaces, $\Sigma_S$ and $\Sigma_N$, as indicated in~\eqref{fixed}. Each of these two surfaces contains the worldline of one static observer and they both have the topology of a 2-sphere in the Euclidean geometry. Moreover, they are by construction minimal surfaces in the sense that their area functional must be coupled to the Einstein--Hilbert action in order to have a well defined variational principle; cf. Equation~\eqref{Itotal}.

By introducing an orbifold parameter $q>1$, we have proposed the existence of an intrinsic field theoretic description of each of the minimal surfaces in terms of a two-dimensional conformal field theory. To this end, we have built up an effective two-dimensional action functional with support on $\Sigma_S$ and $\Sigma_N$, which comprises a free energy inflow coming from dimensionally reducing the four-dimensional Einstein--Hilbert action, plus the corresponding Nambu--Goto term of the surface. The resulting effective action, given in Equation~\eqref{IeffN}, corresponds to that of a Liouville theory on a 2-sphere with a fixed vacuum expectation value of the Liouville field. 

The correspondence between the reduced action on the minimal surfaces and the Liouville theory action provides a non-trivial link between the couplings and parameters of both theories. These consistency conditions, displayed in~\eqref{Pg} and~\eqref{gq}, in particular lead to the $q$-dependent central charge~\eqref{cq}. 
Making use of the thermal Cardy formula, we have computed the Cardy entropy~\eqref{CardyE} which, upon identifying the modular parameter with the inverse of the (dimensionless) temperature $q=T^{-1}$, gives a modular free energy whose modular entropy equals the Gibbons--Hawking entropy. 

The above construction permits the interpretation of the Gibbons--Hawking entropy as representing microscopic degrees of freedom of the massive observer: The back-reaction of such observer induces a conical defect which in turn is the locus of codimension two minimal surface. This two-dimensional surface encode their own field theoretic description in terms of Liouville theory, which yields the central charge~\eqref{cq} and that we propose captures the degrees of freedom of the observer (that are only visible when $q>1$).

We finally studied the $q\to \infty$ limit of the quotient dS$_4/\mathbb Z_q$, which is equivalent to the zero radius limit of the $S^2/\mathbb Z_q$ spindle (see Fig.~\hyperlink{Fig:2}2). In this limit, the four-dimensional geometry reduces to the global geometry of dS$_3$ spacetime where the two minimal surfaces $\Sigma_S$ and $\Sigma_N$ are mapped, upon double analytical continuation, to the future and past conformal boundaries $\mathcal I^+$ and $\mathcal I^-$ of dS$_3$, as indicated in~\eqref{dS3limit}. 

From the relation between the modular parameter and the temperature $q=T^{-1}$, it follows that the limit $q\to\infty$ is also equivalent to zero temperature limit of the Liouville theory on the minimal surfaces.  As a result, the future and past infinities of dS$_3$ inherit from the minimal surfaces a non-thermal Liouville theory. 
Schematically, our findings can be summarized in Figure~\hyperlink{Fig:3}3 below.

\bigskip
\begin{center}
\begin{tikzpicture}
\node at (0,0) {\framebox{dS$_4/\mathbb Z_q$}};
\node at (-5,0) {\framebox{{\footnotesize Global} dS$_3$}};
\node at (5,0) {\framebox{{\footnotesize Static} dS$_4$}};
\node at (-2.5,0.3) {\color{darkred}\footnotesize $q \to \infty$};
\draw [->] (-1.5,0)--(-3.5,0);
\node at (2.5,0.3) {\color{darkred}\footnotesize $q \to 1$};
\draw [->] (1.5,0)--(3.5,0);
\node at (0,-3){
\framebox{\footnotesize
\begin{tabular}{c} 
Liouville theory \\on $\Sigma_{N, S}$   
\end{tabular}}
};
\node at (-5.4,-3){
\framebox{\footnotesize
\begin{tabular}{c} 
Non-thermal \\ Liouville Theory on $\mathcal I^\pm$   
\end{tabular}}
};
\draw [-,thick] (-0.141,-2) arc (-180:0:2pt);
\draw[->,thick](0,-2)--(0,-0.5);
\node at (0,-1.3) {\scriptsize Defects~~(cod-2)};
\draw [-,thick] (-5.141,-2) arc (-180:0:2pt);
\draw[->,thick](-5,-2)--(-5,-0.5);
\node at (-5.25,-1.3) {\scriptsize Boundaries~~(cod-1)};
\node at (-2.5,-2.7) {\color{darkred}\footnotesize $q \to \infty$};
\draw [->] (-2,-3)--(-3,-3);
\node[text width=16cm, text justified] at (-0.3,-4.8){
\small{\hypertarget{Fig:3}\bf Fig.~3}: 
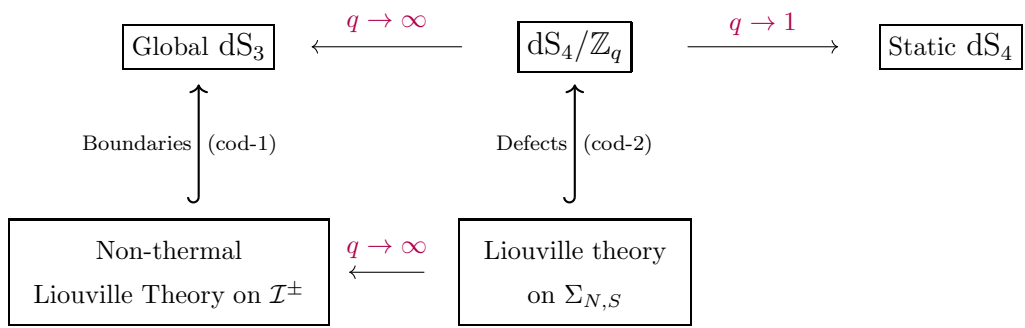
\sffamily{Different limits of the dS$_4/\mathbb Z_q$ geometry and its defect/boundary field theory description.
}};
\end{tikzpicture}
\end{center}

Accordingly, the total central charge of the composite dS$_3$ boundary $\mathcal I^+\cup\mathcal I^-$ comprises two separate contributions, one from $\Sigma_S$ and another one $\Sigma_N$, as displayed in~\eqref{cinfty}.
This can be directly computed by taking the large $q$ limit of the Liouville central charge~\eqref{cq}. The result correctly reproduces the value of the dS$_3$/CFT$_2$ central charge for the boundary field theory.

Regarding directions for future work, one may speculate that our construction belongs to a broader scheme whereby (higher spin) gravity theories are formulated as quasi-topological field theories of the AKSZ type~\cite{Alexandrov:1995kv}. These theories are naturally formulated on manifolds with multiple boundaries and they incorporate extended objects of various codimensions; Hilbert spaces are assigned to boundaries (encoding boundary states of the bulk theory) as well as to defects (encoding defect states labeled by the codimension number). In this moduli space, it is natural to expect that the Hilbert spaces associated to boundaries and defects are related via a (co)dimensional ladder of dualities involving different limits of the moduli parameters. The case presented here would then be a concrete example of such a duality in which the Hilbert space of a codimension two defect in four dimension gives rise, in the large $q$ limit, to the boundary Hilbert space of ${\rm dS}_3$.
We plan to refine and present these ideas in a separate work.

%~~~~~~~~~~~~~~~~~~~~~~~~~~~
\section*{{\large Acknowledgements}}
%~~~~~~~~~~~~~~~~~~~~~~~~~~~
{\sc Ca} would like to thank the hospitality of the Riemann Center for Geometry and Physics at LUH during the completion of this project, where his work was partially supported by a Riemann Fellowship.
{\sc Fd} would like to thank the hospitality of Lebedev Physical Institute during the final stage of this project. 
{\sc Fd} is a Universidad Andres Bello ({\sc Unab) PhD} Scholarship holder, and his work is supported by the Direcci\'on General de 
Investigaci\'on ({\sc Dgi-Unab}).
The work of {\sc Ro} is partially supported by {\sc Fondecyt} Regular grant N$^{\rm o}$ 1170765 and {\sc Conicyt} grant {\sc Dpi} 2014-0115.
The work of {\sc Ps} is in part supported by {\sc Fondecyt} Regular grant N$^{\rm o}$1151107.

\appendix
%~~~~~~~~~~~~~~~~~~~~~~~~~~~
\section{Liouville theory}\label{App:A}
%~~~~~~~~~~~~~~~~~~~~~~~~~~~
In this appendix we collect the most relevant results of Liouville field theory and its 
semiclassical limit. For a more detailed analysis see, for instance, \cite{Distler:1988jt, Seiberg:1990eb, Ginsparg:1993is} and references therein.

\paragraph{Quantum theory.}
Let $(\Sigma, h)$ be a two-dimensional Riemann surface. Liouville theory is an exact two-dimensional conformal field theory on $\Sigma$, defined by the action  
\begin{equation}\label{Qaction}
I_L= \frac{1}{4\pi} \int_\Sigma d^2y \sqrt h 
\Big(h^{ij}\partial_i\Phi \partial_j\Phi + Q\mathcal R \Phi 
+4\pi \mu e^{2\gamma\Phi}\Big)~.
\end{equation}
where the interaction parameter $\mu$ depends on the 
curvature of $\Sigma$, and the coupling $\gamma^2\sim \hbar$ 
controls the quantum effects. When considering the theory
on a Lorentzian manifold, the action~\eqref{Qaction} acquires 
an extra overall minus sign.

Conformal invariance at the full 
quantum level sets the brackground charge
\begin{equation}\label{Q}
Q=\frac{1}{\gamma} + \gamma~,
\end{equation}
which is thus invariant under the shift 
$\gamma\rightarrow\gamma^{-1}$.
In complex coordinates, the (holomorphic part of the)
stress-enery tensor
\begin{equation}
T:= T_{zz} = Q\partial^2 \Phi - (\partial \Phi) ^2 ~,
\end{equation}
gives rise, via the operator product expansion 
\begin{equation}
T(z_1)T(z_2) = \frac{c/2}{(z_1-z_2)^4}
+\frac{2T(z_2)}{(z_1-z_2)^2}
+\frac{\partial T(z_2)}{z_1-z_2} + \cdots ~, 
\end{equation}
to the central charge of the theory
\begin{equation}\label{cL}
c= 1 + 6 Q^2~.
\end{equation}
The vertex operators
\begin{equation}
V_{\alpha}(z) = e^{2\alpha \Phi(z)}~,
\end{equation}
labeled by the quantum number $\alpha$, create the 
spectrum of primary operators of the theory. It follows that 
the operator product
\begin{equation}
T(z_1) V(z_2) = \frac{\Delta\, V_{\alpha}(z_2)}{(z_1-z_2)^2} 
+ \frac{\partial V_{\alpha}(z_2)}{z_1-z_2} + \cdots
\end{equation}
determines the conformal dimension of primaries in terms 
of the background charge and the momentum
\begin{equation}
\Delta = \alpha(Q-\alpha)~.
\end{equation}
The momenta of normalizable states 
\begin{equation}
\alpha = \frac{1}{2} Q + i\lambda ~, \quad 
\lambda\in\mathbb R~,
\end{equation}
in terms of which
\begin{equation}\label{hL}
\Delta=\frac{1}{4}Q^2 + \lambda^2  \geq 
\frac{1}{4}Q^2=:\Delta_0 ~.
\end{equation}
%~~~~~~~~~~~~~~~~~~~~~~~
\paragraph{Semiclassical limit.}
%~~~~~~~~~~~~~~~~~~~~~~~
The semiclassical limit of the theory
is taken through the double scaling 
\begin{equation}
\Phi \rightarrow \gamma^{-1} \Phi ~, \quad
\mu\rightarrow \gamma^{-2} \mu~,
\end{equation}
under which the quantum action~\eqref{Qaction}
scales to 
\begin{equation}\label{SCaction}
I_L= \frac{1}{4\pi \gamma^2} \int_\Sigma d^2y \sqrt h 
\Big(h^{ij}\partial_i\Phi \partial_j\Phi + \mathcal R \Phi 
+4\pi \mu e^{2\Phi}\Big)~,
\end{equation}
in the limit where $\gamma^2\rightarrow0$.
In this regime, the central charge~\eqref{cL} 
and the bound state conformal weight in~\eqref{hL} 
is well approximated by
\begin{equation}\label{cD}
c\approx\frac{6}{\gamma^2}~,\quad
\Delta_0 \approx \frac{c}{24}~.
\end{equation}

%%%%%%%%%%%%%%
\bibliographystyle{JHEP.bst}
{\small\bibliography{LDefects}}

\providecommand{\href}[2]{#2}\begingroup\raggedright\begin{thebibliography}{10}

\bibitem{Arias:2019pzy}
C.~Arias, F.~Diaz and P.~Sundell, \emph{{De Sitter Space and Entanglement}},
  \href{https://doi.org/10.1088/1361-6382/ab5b78}{\emph{Class. Quant. Grav.}
  {\bfseries 37} (2020) 015009},
  [\href{https://arxiv.org/abs/1901.04554}{{\ttfamily 1901.04554}}].

\bibitem{Gibbons:1977mu}
G.~W. Gibbons and S.~W. Hawking, \emph{{Cosmological Event Horizons,
  Thermodynamics, and Particle Creation}},
  \href{https://doi.org/10.1103/PhysRevD.15.2738}{\emph{Phys. Rev.} {\bfseries
  D15} (1977) 2738--2751}.

\bibitem{Deser:1983nh}
S.~Deser and R.~Jackiw, \emph{{Three-Dimensional Cosmological Gravity: Dynamics
  of Constant Curvature}},
  \href{https://doi.org/10.1016/0003-4916(84)90025-3}{\emph{Annals Phys.}
  {\bfseries 153} (1984) 405--416}.

\bibitem{Deser:1983tn}
S.~Deser, R.~Jackiw and G.~'t~Hooft, \emph{{Three-Dimensional Einstein Gravity:
  Dynamics of Flat Space}},
  \href{https://doi.org/10.1016/0003-4916(84)90085-X}{\emph{Annals Phys.}
  {\bfseries 152} (1984) 220}.

\bibitem{Deser:1989cf}
S.~Deser and R.~Jackiw, \emph{{String Sources in (2+1)-dimensional Gravity}},
  \href{https://doi.org/10.1016/0003-4916(89)90140-1}{\emph{Annals Phys.}
  {\bfseries 192} (1989) 352}.

\bibitem{Park:1998qk}
M.-I. Park, \emph{{Statistical entropy of three-dimensional Kerr-de Sitter
  space}}, \href{https://doi.org/10.1016/S0370-2693(98)01119-8}{\emph{Phys.
  Lett.} {\bfseries B440} (1998) 275--282},
  [\href{https://arxiv.org/abs/hep-th/9806119}{{\ttfamily hep-th/9806119}}].

\bibitem{Banados:1998tb}
M.~Banados, T.~Brotz and M.~E. Ortiz, \emph{{Quantum three-dimensional de
  Sitter space}}, \href{https://doi.org/10.1103/PhysRevD.59.046002}{\emph{Phys.
  Rev.} {\bfseries D59} (1999) 046002},
  [\href{https://arxiv.org/abs/hep-th/9807216}{{\ttfamily hep-th/9807216}}].

\bibitem{Strominger:2001pn}
A.~Strominger, \emph{{The dS / CFT correspondence}},
  \href{https://doi.org/10.1088/1126-6708/2001/10/034}{\emph{JHEP} {\bfseries
  10} (2001) 034}, [\href{https://arxiv.org/abs/hep-th/0106113}{{\ttfamily
  hep-th/0106113}}].

\bibitem{Nojiri:2001mf}
S.~Nojiri and S.~D. Odintsov, \emph{{Conformal anomaly from dS / CFT
  correspondence}},
  \href{https://doi.org/10.1016/S0370-2693(01)00869-3}{\emph{Phys. Lett.}
  {\bfseries B519} (2001) 145--148},
  [\href{https://arxiv.org/abs/hep-th/0106191}{{\ttfamily hep-th/0106191}}].

\bibitem{Klemm:2001ea}
D.~Klemm, \emph{{Some aspects of the de Sitter / CFT correspondence}},
  \href{https://doi.org/10.1016/S0550-3213(02)00007-X}{\emph{Nucl. Phys.}
  {\bfseries B625} (2002) 295--311},
  [\href{https://arxiv.org/abs/hep-th/0106247}{{\ttfamily hep-th/0106247}}].

\bibitem{Cacciatori:2001un}
S.~Cacciatori and D.~Klemm, \emph{{The Asymptotic dynamics of de Sitter gravity
  in three-dimensions}},
  \href{https://doi.org/10.1088/0264-9381/19/3/312}{\emph{Class. Quant. Grav.}
  {\bfseries 19} (2002) 579--588},
  [\href{https://arxiv.org/abs/hep-th/0110031}{{\ttfamily hep-th/0110031}}].

\bibitem{Klemm:2002ir}
D.~Klemm and L.~Vanzo, \emph{{De Sitter gravity and Liouville theory}},
  \href{https://doi.org/10.1088/1126-6708/2002/04/030}{\emph{JHEP} {\bfseries
  04} (2002) 030}, [\href{https://arxiv.org/abs/hep-th/0203268}{{\ttfamily
  hep-th/0203268}}].

\bibitem{thurston2014three}
W.~Thurston and S.~Levy, \emph{Three-Dimensional Geometry and Topology}.
\newblock No.~v. 1 in Princeton Mathematical Series. Princeton University
  Press, 2014.

\bibitem{Fursaev:1995ef}
D.~V. Fursaev and S.~N. Solodukhin, \emph{{On the description of the Riemannian
  geometry in the presence of conical defects}},
  \href{https://doi.org/10.1103/PhysRevD.52.2133}{\emph{Phys. Rev.} {\bfseries
  D52} (1995) 2133--2143},
  [\href{https://arxiv.org/abs/hep-th/9501127}{{\ttfamily hep-th/9501127}}].

\bibitem{Polyakov:1981rd}
A.~M. Polyakov, \emph{{Quantum Geometry of Bosonic Strings}},
  \href{https://doi.org/10.1016/0370-2693(81)90743-7}{\emph{Phys. Lett.}
  {\bfseries B103} (1981) 207--210}.

\bibitem{Solodukhin:1998tc}
S.~N. Solodukhin, \emph{{Conformal description of horizon's states}},
  \href{https://doi.org/10.1016/S0370-2693(99)00398-6}{\emph{Phys. Lett.}
  {\bfseries B454} (1999) 213--222},
  [\href{https://arxiv.org/abs/hep-th/9812056}{{\ttfamily hep-th/9812056}}].

\bibitem{Jackiw:2005su}
R.~Jackiw, \emph{{Weyl symmetry and the Liouville theory}},
  \href{https://doi.org/10.1007/s11232-006-0090-9}{\emph{Theor. Math. Phys.}
  {\bfseries 148} (2006) 941--947},
  [\href{https://arxiv.org/abs/hep-th/0511065}{{\ttfamily hep-th/0511065}}].

\bibitem{Balasubramanian:1999zv}
V.~Balasubramanian and S.~F. Ross, \emph{{Holographic particle detection}},
  \href{https://doi.org/10.1103/PhysRevD.61.044007}{\emph{Phys. Rev.}
  {\bfseries D61} (2000) 044007},
  [\href{https://arxiv.org/abs/hep-th/9906226}{{\ttfamily hep-th/9906226}}].

\bibitem{Martinec:2002xq}
E.~J. Martinec and W.~McElgin, \emph{{Exciting AdS orbifolds}},
  \href{https://doi.org/10.1088/1126-6708/2002/10/050}{\emph{JHEP} {\bfseries
  10} (2002) 050}, [\href{https://arxiv.org/abs/hep-th/0206175}{{\ttfamily
  hep-th/0206175}}].

\bibitem{Balasubramanian:2003kq}
V.~Balasubramanian, A.~Naqvi and J.~Simon, \emph{{A Multiboundary AdS orbifold
  and DLCQ holography: A Universal holographic description of extremal black
  hole horizons}},
  \href{https://doi.org/10.1088/1126-6708/2004/08/023}{\emph{JHEP} {\bfseries
  08} (2004) 023}, [\href{https://arxiv.org/abs/hep-th/0311237}{{\ttfamily
  hep-th/0311237}}].

\bibitem{deBoer:2010ac}
J.~de~Boer, M.~M. Sheikh-Jabbari and J.~Simon, \emph{{Near Horizon Limits of
  Massless BTZ and Their CFT Duals}},
  \href{https://doi.org/10.1088/0264-9381/28/17/175012}{\emph{Class. Quant.
  Grav.} {\bfseries 28} (2011) 175012},
  [\href{https://arxiv.org/abs/1011.1897}{{\ttfamily 1011.1897}}].

\bibitem{Chen:2018vkw}
C.-B. Chen, W.-C. Gan, F.-W. Shu and B.~Xiong, \emph{{Quantum information
  metric of conical defect}},
  \href{https://doi.org/10.1103/PhysRevD.98.046008}{\emph{Phys. Rev.}
  {\bfseries D98} (2018) 046008},
  [\href{https://arxiv.org/abs/1804.08358}{{\ttfamily 1804.08358}}].

\bibitem{Cardy:1986ie}
J.~L. Cardy, \emph{{Operator Content of Two-Dimensional Conformally Invariant
  Theories}}, \href{https://doi.org/10.1016/0550-3213(86)90552-3}{\emph{Nucl.
  Phys.} {\bfseries B270} (1986) 186--204}.

\bibitem{Bloete:1986qm}
H.~W.~J. Bloete, J.~L. Cardy and M.~P. Nightingale, \emph{{Conformal
  Invariance, the Central Charge, and Universal Finite Size Amplitudes at
  Criticality}}, \href{https://doi.org/10.1103/PhysRevLett.56.742}{\emph{Phys.
  Rev. Lett.} {\bfseries 56} (1986) 742--745}.

\bibitem{Hartman:2014oaa}
T.~Hartman, C.~A. Keller and B.~Stoica, \emph{{Universal Spectrum of 2d
  Conformal Field Theory in the Large c Limit}},
  \href{https://doi.org/10.1007/JHEP09(2014)118}{\emph{JHEP} {\bfseries 09}
  (2014) 118}, [\href{https://arxiv.org/abs/1405.5137}{{\ttfamily 1405.5137}}].

\bibitem{Laflamme:1988wg}
R.~Laflamme, \emph{{Geometry and Thermofields}},
  \href{https://doi.org/10.1016/0550-3213(89)90191-0}{\emph{Nucl. Phys.}
  {\bfseries B324} (1989) 233--252}.

\bibitem{Barvinsky:1994jca}
A.~O. Barvinsky, V.~P. Frolov and A.~I. Zelnikov, \emph{{Wavefunction of a
  Black Hole and the Dynamical Origin of Entropy}},
  \href{https://doi.org/10.1103/PhysRevD.51.1741}{\emph{Phys. Rev.} {\bfseries
  D51} (1995) 1741--1763},
  [\href{https://arxiv.org/abs/gr-qc/9404036}{{\ttfamily gr-qc/9404036}}].

\bibitem{Jacobson:1994fp}
T.~Jacobson, \emph{{A Note on Hartle-Hawking vacua}},
  \href{https://doi.org/10.1103/PhysRevD.50.R6031}{\emph{Phys. Rev.} {\bfseries
  D50} (1994) R6031--R6032},
  [\href{https://arxiv.org/abs/gr-qc/9407022}{{\ttfamily gr-qc/9407022}}].

\bibitem{Baez:2011}
J.~C. Baez, \emph{{Renyi Entropy and Free Energy}},
  \href{https://arxiv.org/abs/arXiv:1102.2098}{{\ttfamily arXiv:1102.2098}}.

\bibitem{Rangamani:2016dms}
M.~Rangamani and T.~Takayanagi, \emph{{Holographic Entanglement Entropy}},
  \href{https://doi.org/10.1007/978-3-319-52573-0}{\emph{Lect. Notes Phys.}
  {\bfseries 931} (2017) pp.1--246},
  [\href{https://arxiv.org/abs/1609.01287}{{\ttfamily 1609.01287}}].

\bibitem{Calabrese:2004eu}
P.~Calabrese and J.~L. Cardy, \emph{{Entanglement entropy and quantum field
  theory}}, \href{https://doi.org/10.1088/1742-5468/2004/06/P06002}{\emph{J.
  Stat. Mech.} {\bfseries 0406} (2004) P06002},
  [\href{https://arxiv.org/abs/hep-th/0405152}{{\ttfamily hep-th/0405152}}].

\bibitem{Maldacena:1998ab}
J.~M. Maldacena and A.~Strominger, \emph{Statistical entropy of de sitter
  space}, {\emph{JHEP} {\bfseries 9802} (1998) 014},
  [\href{https://arxiv.org/abs/gr-qc/9801096}{{\ttfamily gr-qc/9801096}}].

\bibitem{Alexandrov:1995kv}
M.~Alexandrov, A.~Schwarz, O.~Zaboronsky and M.~Kontsevich, \emph{{The Geometry
  of the master equation and topological quantum field theory}},
  \href{https://doi.org/10.1142/S0217751X97001031}{\emph{Int. J. Mod. Phys.}
  {\bfseries A12} (1997) 1405--1429},
  [\href{https://arxiv.org/abs/hep-th/9502010}{{\ttfamily hep-th/9502010}}].

\bibitem{Distler:1988jt}
J.~Distler and H.~Kawai, \emph{{Conformal Field Theory and 2D Quantum
  Gravity}}, \href{https://doi.org/10.1016/0550-3213(89)90354-4}{\emph{Nucl.
  Phys.} {\bfseries B321} (1989) 509--527}.

\bibitem{Seiberg:1990eb}
N.~Seiberg, \emph{{Notes on quantum Liouville theory and quantum gravity}},
  \href{https://doi.org/10.1143/PTPS.102.319}{\emph{Prog. Theor. Phys. Suppl.}
  {\bfseries 102} (1990) 319--349}.

\bibitem{Ginsparg:1993is}
P.~H. Ginsparg and G.~W. Moore, \emph{{Lectures on 2-D gravity and 2-D string
  theory}},  in \emph{{Proceedings, Theoretical Advanced Study Institute (TASI
  92): From Black Holes and Strings to Particles: Boulder, USA, June 1-26,
  1992}}, pp.~277--469, 1993,
  \href{https://arxiv.org/abs/hep-th/9304011}{{\ttfamily hep-th/9304011}}.

\end{thebibliography}\endgroup
\end{document}